\documentclass[a4paper,11pt]{article}
\usepackage{pos}

\title{Unsupervised machine learning correlations in EoS of neutron stars}
\author*[a,c]{Ronaldo V. Lobato}
\author[b]{Emanuel V. Chimanski}
\author[c]{Carlos A. Bertulani}

\affiliation[a]{Departamento de F\'isica, Universidad de los Andes, Bogot\'a, Colombia.}
\affiliation[b]{Lawrence Livermore National Laboratory, Livermore, CA, USA}
\affiliation[c]{Department of Physics and Astronomy, Texas A\&M
  University - Commerce, Commerce, TX, USA.}

\emailAdd{r.vieira@uniandes.edu.co}
\emailAdd{chimanski1@llnl.gov}
\emailAdd{carlos.bertulani@tamuc.edu}

\abstract{
  Neutron stars are compact objects of large interest in the
  nuclear astrophysics community. The extreme conditions present in
  such systems impose big challenges to our current microscopic
  models of nuclear structure. Equation of states (EoS) are frequently
  derived from sophisticated quantum mechanical models, such as:
  relativistic, non-relativistic and many mean-field approaches. Every single model, in general, contains many parameters such as the NN interaction strength, particle compositions, etc. These are particular features of each model and can be represented by numbers and categories in a machine learning context. Different choices of features will affect EoS properties leading to different macroscopic properties of the star. In this work we analyze a selection of
  EoS containing a variety of different physics models. One of our objectives is to develop tools
  that enable a better understanding of the correlations among
  the different model features and the outcome produced by them when
  employed to model neutron stars.
}

\FullConference{%
  XV International Workshop on Hadron Physics (XV Hadron Physics)
  13 -17 September 2021 \\
  Online, hosted by Instituto Tecnológico de Aeronáutica, São José dos Campos, Brazil\\
  }

\begin{document}
\maketitle

\section{Introduction}\label{intro}
Neutron stars (NS) are supernova remnants with strong gravitational fields and rapid rotation. They
consist of the highest density objects in the Universe with numbers ranging from a few ${\rm g/cm^3}$ at their surface to more than ${\rm 10^{15}\ g/cm^3}$ at their center \cite{haensel/2007}.

The microscopic description of these systems have been extensively studied in the last few decades and yet, a unified and complete theoretical understanding is missing. It is still difficult to exclude the many possible scenarios and converge to a single set of parameterization and constraints on the EoS even when both astrophysical observations and nuclear physics experiments are considered \cite{Burgio-2021}.

Part of the challenge here relates to extreme physical environments, e.g. large matter-energy densities, and the limits of our models that contain parameters adjusted to reproduce, at their best, nuclear properties on natural conditions present on Earth.
These challenges open up space for many new functional parameterization and
models that can reproduce the physics of matter from Earth-like conditions and be extrapolated to the high density stellar environment. In this work, we demonstrate how
simple unsupervised machine learning techniques can help us to
identify important correlations among the various EoS of dense
matter, commonly used to model NS. To understand the outcomes of
different equations of states, we employ dimensional
reduction algorithms such as: Multi Dimensional
Scaling (MDS), Principal Value Component Analysis (PCA), and
t-distributed Stochastic Neighbor Embedding (t-SNE). The dimensional reduction provided by such tools helps to discover underlying structures present in the different physics models. We have selected a set of popular EoS that represent different models to demonstrate our approach. In Section \ref{eos} we give a short description of the physics of equation of state. Section \ref{ml} presents a brief description of the ML methods employed in this study, as well as our results and discussions. Final remarks are provided in Sec. \ref{conc}.

\section{Equation of state}\label{eos}
The high regime/densities of the EoS describing the NS interior has
not been fully constrained, leaving an open question in nuclear
astrophysics. Only a few microscopic physics constraints are
currently possible when one considers neutron stars: electric neutrality, beta equilibrium, positive pressure, $p\geq 0$, and $dp/d\rho > 0$ from the Chateliers' principle and finally causality, i.e., the speed of the sound $v_s$ must be less than the speed of light $c$. The uncertainty in the NS interior leads to a large variety of EoS, roughly distinguished by the compressibility of the nuclear matter (i.e., softness and stiffness, which is associated to the speed of sound in the matter) and its behavior in large energy density regimes.
There are several methods to calculate the EoS: Perturbation expansion within the Brueckner-Bethe-Goldstone-[Hartree-Fock] theory (BHF), perturbation expansion within the Green's-function theory, variational method (VF), energy density functional (EDF) theory, relativistic mean-field (RMF) models~\cite{bethe/1971, ring/1980, blaizot/1985, machleidt/1989, akmal/1997,haensel/2007}. Point-coupling and non-relativistic models such as Skyrme and Gogny Hartree-Fock Bogoliubov (HFB) theories are also used \cite{nikolaus/1992, friar/1996, skyrme/1958, bell/1956, skyrme/2006, decharge/1980, berger/1991}. Skyrme and RMF models span more than 500 parameterization possibilities, which raises doubt whether they can reproduce different density environments simultaneously. Some studies try to constrain all these parameters in the
vicinity of the nuclear saturation density~\cite{dutra/2012, dutra/2014,lourenco/2019}, while others consider the binary neutron star merger observation released by
LIGO-VIRGO collaborations \cite{abbott/2017, abbott/2017a, abbott/2017b} as constrain for the mass-radius of NS and consequently for the EoS which generates them \cite{radice/2018a, motta/2019, margalit/2017, bauswein/2017, gamba/2019,lourenco/2020, essick/2020}.

One can expect correlations among all the different existing models,
but their variety and large set of parameters involved poses barriers
for a more detailed study. Here we selected a small set of EoS derived from different physics models, the most commonly used in the
literature \cite{lattimer/2001, lackey/2006, bejger/2005, ozel/2016}. We summarize their characteristics in Table \ref{tab:EoS}. Models with same composition but with non-relativistic and relativistic approaches are included.

\begin{table}[ht]
\centering
\begin{tabular}[t]{lllll}
\hline
EoS	&	Composition	&	Model	&	Approach	&	Potential	\\
\hline
\hline
APR1-4	\cite{Akmal-1998} &	npem & non-relativistic &	Variational	&	Two-three body	\\
BBB2	\cite{Baldo-1997} &	npem & non-relativistic	&	BHF	&	Two-three body			\\
FPS	    \cite{Friedman-1981} & npem	& non-relativistic & Variational	&	Two-three body	\\
SLy4	    \cite{Douchin-2001} &	npem		&	non-relativistic	&	EDF	&	Two-body			\\
PAL6    \cite{Prakash-1988} &	npem		&	non-relativistic	&	Schematic potential	&   Two-body \\
WWF1-3	    \cite{Wiringa-1988} &	npem		&	non-relativistic	&	Variational	&	Two-body	\\
ENG	    \cite{Engvik-1996} &	npem		&	relativistic	&	Dirac-BHF	&	Meson exchange	\\
MPA1    \cite{Muther-1987} &	npem		&	relativistic	&	Dirac-BHF	&	Meson exchange	\\
MS1-2,1b\cite{Muller-1996} &	npem		&	relativistic	&	MF	&	Meson exchange			\\
BPAL12  \cite{Zuo-1999}	&	npem	&	relativistic	&	Dirac-BHF	&	Two-body			\\
PS      \cite{Pandharipande-1975}	&	meson		&	non-relativistic	&	Potential	&	Two-body \\
GS1-2   \cite{Glendenning-1999}	&	meson	&		relativistic	&	MF	&	Meson exchange                  \\
GNH3    \cite{Glendenning-1985}	&	meson	&		relativistic	&	MF	&	Two-body			\\
H1-7    \cite{Lackey-2006}	&	hyperon	&	relativistic		&	MF	&	Meson exchange                  \\
PCL2    \cite{Prakash-1995}	&	hyperon	&	relativistic		&	MF	&	Meson exchange                  \\
ALF1-4  \cite{Alford-2005}	&	quark	&	relativistic	&	MIT	&	Gluons (QCD)			\\
\hline
\end{tabular}
\caption{\label{tab:EoS}Summary of selected EoS. The
  composition npem stands for nucleonic matter in $\beta$-equilibrium.
  Meson, hyperon and quarks are models collective known as K/$\pi$/H/q
models. }
\end{table}

\section{Correlations and Data Grouping}\label{ml}

The first unsupervised machine learning method that we have used is the Principal
Components Analysis (PCA) \cite{jolliffe/2013}. This algorithm is part of the {\it dimension reduction}. It transforms the characteristics of a dataset into a new set of features called Principal Components. By doing this, many variables across
the full dataset are effectively compressed in fewer feature
columns. This reduction creates a new set of 'uncorrelated' variables as functions of the old features. We use this approach in a multidimensional scaling (MDS) \cite{cox/2008} technique to find a low-dimensional graphical projection, a 2D data in our case. The resulting reduced data is presented in a grouped form rule by their best similarities obtained with data point distances. The second method used in this work is the t-distributed
Stochastic Neighbor Embedding (t-SNE) \cite{NIPS2002_6150ccc6}. In this case, we again reduce the dimensions of the data, trying at the same time to keep similar (EoS) distances close and dissimilar instances apart. Unlike PCA, which is a linear technique, t-SNE is nonlinear, and it permits to separate data that cannot be separated by any straight line.

\subsection{Using the MDS}

\begin{figure}[h!]
\centering
\includegraphics[scale = 0.360]{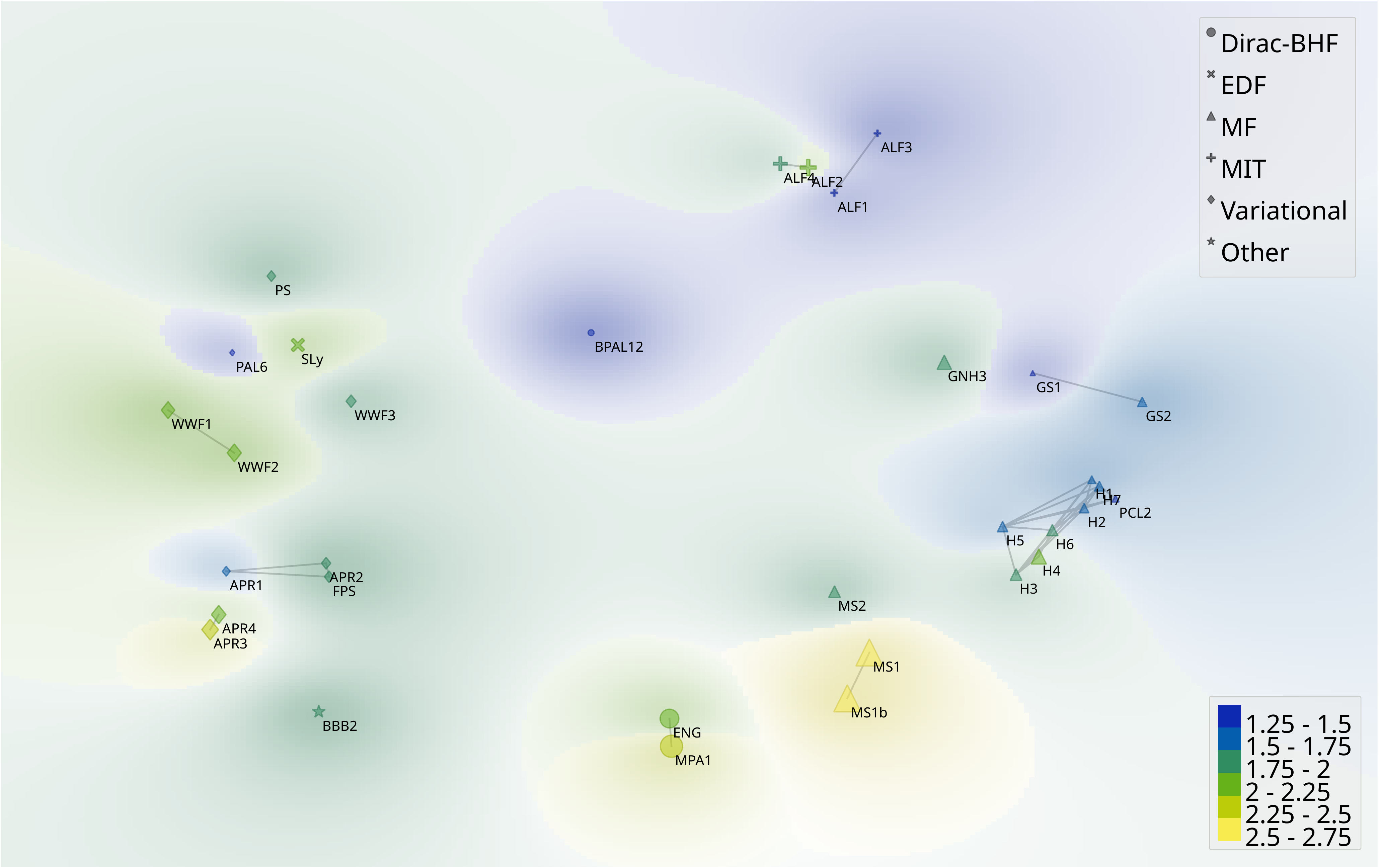}
\caption{\label{multi1} Multidimensional scaling projection using PCA. We use $M_{\rm max}$ as colors. The different geometric forms represent the feature approach utilized to generate the EoS.}
\end{figure}

In figure \ref{multi1} we have a multidimensional scaling projection
considering the PCA reduction in our dataset from the EoS table \ref{tab:EoS}. In each denser color, we display the maximum masses for the EoS
sample: in blue we have the less massive stars and in yellow the most
massive ones. In geometric forms, we have the approach utilized to
generate the EoS. We can see the formation of clusters, the first
pattern is due to the parameterization within each EoS, i.e., one can see a cluster formation in ALF1-4, H1-7, APR3-4, APR1-2; which is obvious at a first glance. However, what is remarkable is that the EoS
responsible for the highest/lowest mass are scattered: One cannot see
a clusterization due this physical quantity because there is no clear-colored region.

\begin{figure}[h!]
\centering
\includegraphics[scale = 0.360]{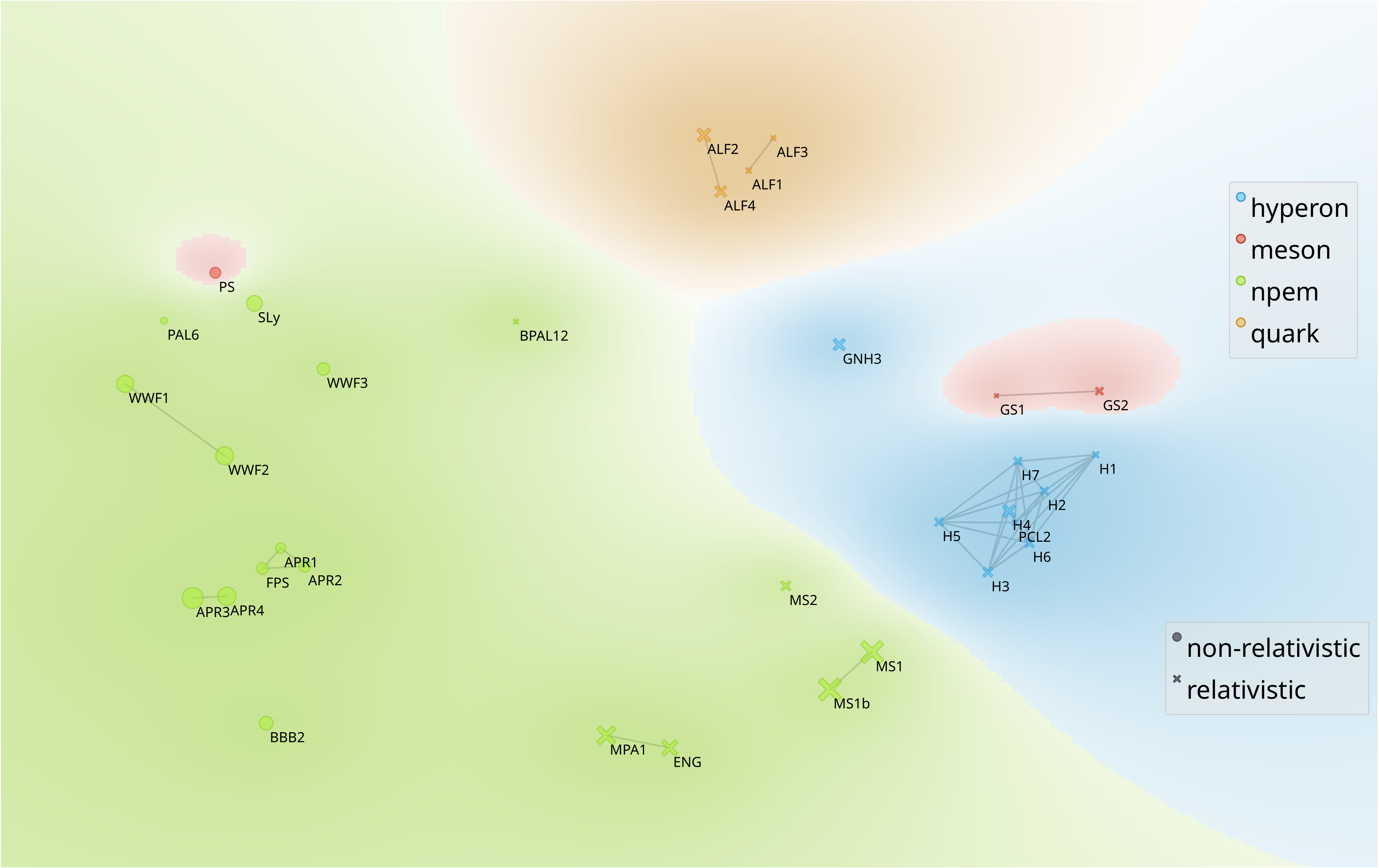}
\caption{\label{multi2} Multidimensional scaling projection using
  PCA. We use the composition of the EoS as colors. In geometric forms, we display if the model is non-relativistic or relativistic.}
\end{figure}

In figure \ref{multi2} we have again the MDS considering PCA reduction. In this second case, we have in colors the composition of the EoS and in geometric forms if the model is relativistic or not. We have the
formation of clusters as the previous case, however here it is
possible to see a separation of the colored regions due to the
composition of the EoS. One sees one exception, PS, and one can
consider it as an outlier in terms of composition. However, there is a correlation with
PAL6 and SLy4 which overrules the composition of the EoS as principal characteristic. It is also
possible to see a separation among the
non-relativistic and relativistic models. On the left side of the
graphic we have non-relativistic nucleonic EoS and from the
bottom-center, where one finds a group of nucleonic, to
the upper-right side, the relativistic ones, where the K/$\pi$/H/q
models are located. Regarding the relativistic-K/$\pi$/H/q
models, it is possible to see one EoS that is an
outlier, the GS with its two parameterizations. In figure
\ref{multi2}, we have a clear separation of the relativistic and
non-relativistic models, however comparing with the previous
Fig. \ref{multi1}, we see no correlation of the maximum mass with the
nature of the models.

\subsection{Using the t-SNE}

\begin{figure}[h!]
\centering
\includegraphics[scale = 0.360]{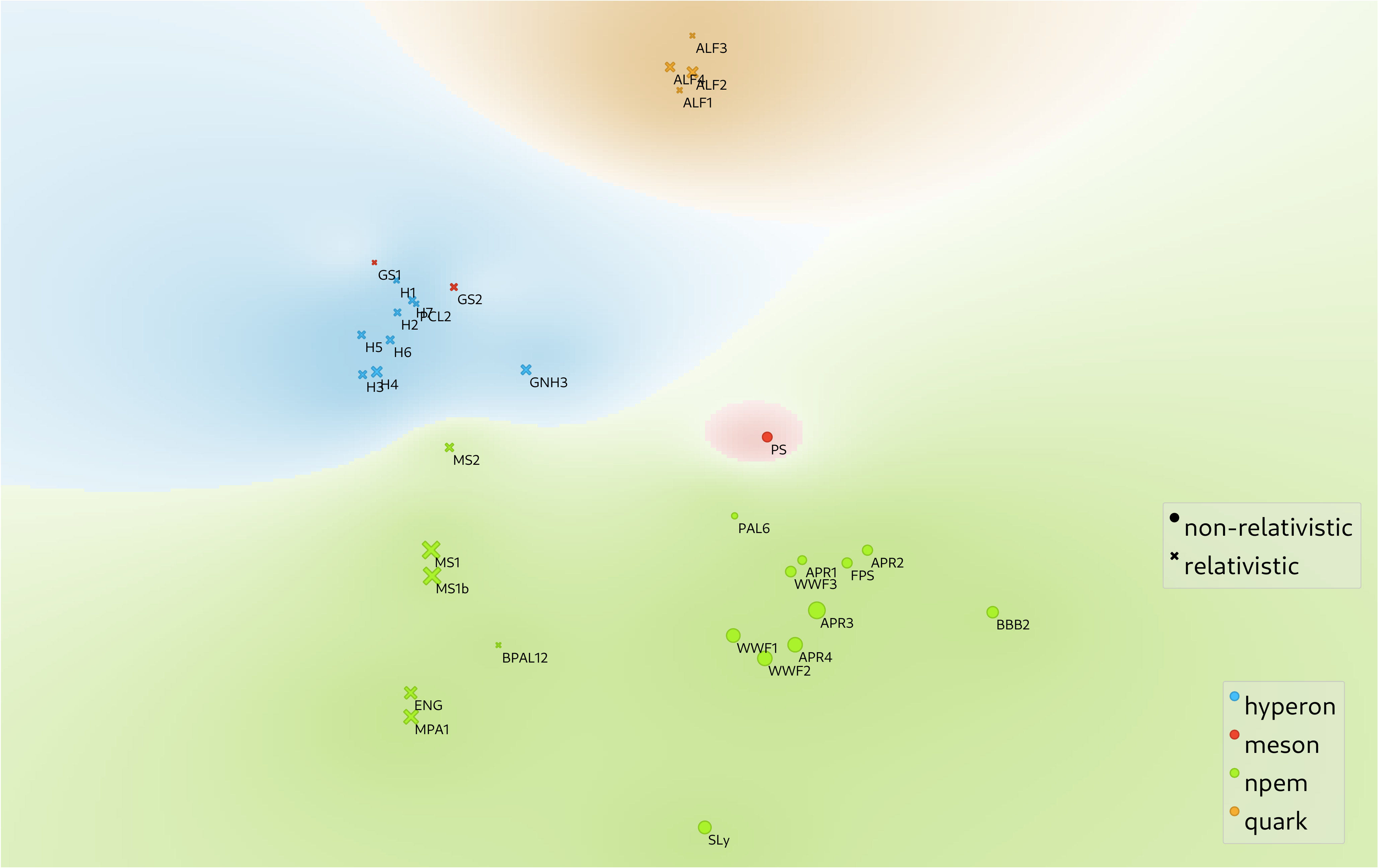}
\caption{\label{sne1} t-SNE using PCA with 6 components. We use the
  composition of the EoS as colors. In geometric forms, we have the different models.}
\end{figure}

In figure \ref{sne1} it is shown the correlation of the dataset using
the t-SNE which used a PCA with 6 principal components. In this figure,
we have again the formation of well-defined regions according to the
composition of the EoS. Again, PS is an outlier; however, this time it is
less correlated with the hadronic models in comparison with
Fig. \ref{multi2}. Using t-SNE we can see that the GS EoS is more
related to the K/$\pi$/H/q models and that they are more strongly
correlated, i.e., ones can see that the nucleonic models are closer
as well as the K/$\pi$/H/q models. One needs a strong jittery to
separate the EoS.

\section{Concluding Remarks}\label{conc}
In this work, we have investigated correlations in a small sample of EoS. We made use of different unsupervised
machine learning algorithms in a dimensionality reduction approach. Using three
algorithms (PCA, MDS and t-SNE) we were able to visualize correlations, i.e., features which are hidden when dealing with complex physical
models. This is one of the key points when
employing machine learning techniques, and can be used to provide
feedback for the theoretical models. This approach can be very helpful to achieve a better description of nuclear matter at high densities and temperatures.

We want to stress that a large set of EoS and its features, as well
as combination of precise data from advanced detectors (VIRGO-LIGO-KAGRA and eXTP) is required for best use of machine learning models. We believe that the use of unsupervised ML approaches and sophisticated visualization tools can help to categorize the many models available in the market. Refining the microscopic models we can find a more realistic EoS that can describe the internal structure of neutron stars.

\section*{Acknowledgments}
RVL and CAB have been partly funded by U.S. Department of Energy (DOE) under grant DE--FG02--08ER4153 and (RVL) by UNIANDES University. This work was performed in part under the auspices of the U.S. Department of Energy by Lawrence Livermore National Laboratory under Contract DE-AC52-07NA27344.

\bibliographystyle{JHEP}
\bibliography{biblio}

\providecommand{\href}[2]{#2}\begingroup\raggedright\begin{thebibliography}{10}

\bibitem{haensel/2007}
P.~Haensel, A.Y.~Potekhin and D.G.~Yakovlev, \emph{Neutron {{Stars}} 1:
  Equation of {{State}} and {{Structure}}}, Astrophysics and {{Space Science
  Library}}, {{Neutron Stars}}, {Springer-Verlag}, {New York} (2007),
  \href{https://doi.org/10.1007/978-0-387-47301-7}{10.1007/978-0-387-47301-7}.

\bibitem{Burgio-2021}
G.~Burgio, H.-J.~Schulze, I.~Vidaña and J.-B.~Wei, \emph{Neutron stars and the
  nuclear equation of state},
  \href{https://doi.org/https://doi.org/10.1016/j.ppnp.2021.103879}{\emph{Progress
  in Particle and Nuclear Physics} {\bfseries 120} (2021) 103879}.

\bibitem{bethe/1971}
H.A.~Bethe, \emph{Theory of {{Nuclear Matter}}},
  \href{https://doi.org/10.1146/annurev.ns.21.120171.000521}{\emph{Annual
  Review of Nuclear Science} {\bfseries 21} (1971) 93}.

\bibitem{ring/1980}
P.~Ring and P.~Schuck, \emph{The {{Nuclear Many}}-{{Body Problem}}},
  Theoretical and {{Mathematical Physics}}, {{The Nuclear Many}}-{{Body
  Problem}}, {Springer-Verlag}, {Berlin Heidelberg} (1980).

\bibitem{blaizot/1985}
J.-P.~Blaizot and G.~Ripka, \emph{Quantum {{Theory}} of {{Finite Systems}}},
  {The MIT Press}, {Cambridge, Mass} (Dec., 1985).

\bibitem{machleidt/1989}
R.~Machleidt, \emph{The {{Meson Theory}} of {{Nuclear Forces}} and {{Nuclear
  Structure}}},  in \emph{Advances in {{Nuclear Physics}}}, J.W.~Negele and
  E.~Vogt, eds., Advances in {{Nuclear Physics}}, ({Boston, MA}), pp.~189--376,
  {Springer US} (1989),
  \href{https://doi.org/10.1007/978-1-4613-9907-0_2}{DOI}.

\bibitem{akmal/1997}
A.~Akmal and V.R.~Pandharipande, \emph{Spin-isospin structure and pion
  condensation in nucleon matter},
  \href{https://doi.org/10.1103/physrevc.56.2261}{\emph{Physical Review C}
  {\bfseries 56} (1997) 2261}.

\bibitem{nikolaus/1992}
B.A.~Nikolaus, T.~Hoch and D.G.~Madland, \emph{Nuclear ground state properties
  in a relativistic point coupling model},
  \href{https://doi.org/10.1103/physrevc.46.1757}{\emph{Physical Review C}
  {\bfseries 46} (1992) 1757}.

\bibitem{friar/1996}
J.L.~Friar, D.G.~Madland and B.W.~Lynn, \emph{{{QCD}} scales in finite nuclei},
  \href{https://doi.org/10.1103/physrevc.53.3085}{\emph{Physical Review C}
  {\bfseries 53} (1996) 3085}.

\bibitem{skyrme/1958}
T.H.R.~Skyrme, \emph{The effective nuclear potential},
  \href{https://doi.org/10.1016/0029-5582(58)90345-6}{\emph{Nuclear Physics}
  {\bfseries 9} (1958) 615}.

\bibitem{bell/1956}
J.S.~Bell and T.H.R.~Skyrme, \emph{{{CVIII}}. {{The}} nuclear spin-orbit
  coupling}, \href{https://doi.org/10.1080/14786435608238187}{\emph{The
  Philosophical Magazine: A Journal of Theoretical Experimental and Applied
  Physics} {\bfseries 1} (1956) 1055}.

\bibitem{skyrme/2006}
T.H.R.~Skyrme, \emph{{{CVII}}. {{The}} nuclear surface},
  \href{https://doi.org/10.1080/14786435608238186}{\emph{Philosophical
  Magazine} (2006) }.

\bibitem{decharge/1980}
J.~Decharg{\'e} and D.~Gogny, \emph{Hartree-{{Fock}}-{{Bogolyubov}}
  calculations with the \${{D1}}\$ effective interaction on spherical nuclei},
  \href{https://doi.org/10.1103/physrevc.21.1568}{\emph{Physical Review C}
  {\bfseries 21} (1980) 1568}.

\bibitem{berger/1991}
J.F.~Berger, M.~Girod and D.~Gogny, \emph{Time-dependent quantum collective
  dynamics applied to nuclear fission},
  \href{https://doi.org/10.1016/0010-4655(91)90263-k}{\emph{Computer Physics
  Communications} {\bfseries 63} (1991) 365}.

\bibitem{dutra/2012}
M.~Dutra, O.~Louren{\c c}o, J.S.~S{\'a}~Martins, A.~Delfino, J.R.~Stone and
  P.D.~Stevenson, \emph{Skyrme interaction and nuclear matter constraints},
  \href{https://doi.org/10.1103/physrevc.85.035201}{\emph{Physical Review C}
  {\bfseries 85} (2012) 035201}.

\bibitem{dutra/2014}
M.~Dutra, O.~Louren{\c c}o, S.S.~Avancini, B.V.~Carlson, A.~Delfino,
  D.P.~Menezes et~al., \emph{Relativistic mean-field hadronic models under
  nuclear matter constraints},
  \href{https://doi.org/10.1103/physrevc.90.055203}{\emph{Physical Review C}
  {\bfseries 90} (2014) 055203}.

\bibitem{lourenco/2019}
O.~Louren{\c c}o, M.~Dutra, C.H.~Lenzi, C.V.~Flores and D.P.~Menezes,
  \emph{Consistent relativistic mean-field models constrained by {{GW170817}}},
  \href{https://doi.org/10.1103/physrevc.99.045202}{\emph{Physical Review C}
  {\bfseries 99} (2019) 045202}.

\bibitem{abbott/2017}
B.P.~Abbott, R.~Abbott, T.D.~Abbott, F.~Acernese, K.~Ackley, C.~Adams et~al.,
  \emph{Gravitational {{Waves}} and {{Gamma}}-{{Rays}} from a {{Binary Neutron
  Star Merger}}: {{GW170817}} and {{GRB 170817A}}},
  \href{https://doi.org/10.3847/2041-8213/aa920c}{\emph{The Astrophysical
  Journal} (2017) } [\href{https://arxiv.org/abs/1710.05834}{{\ttfamily
  1710.05834}}].

\bibitem{abbott/2017a}
B.P.~Abbott, R.~Abbott, T.D.~Abbott, F.~Acernese, K.~Ackley, C.~Adams et~al.,
  \emph{{{GW170817}}: Observation of {{Gravitational Waves}} from a {{Binary
  Neutron Star Inspiral}}},
  \href{https://doi.org/10.1103/physrevlett.119.161101}{\emph{Physical Review
  Letters} {\bfseries 119} (2017) 161101}
  [\href{https://arxiv.org/abs/1710.05836}{{\ttfamily 1710.05836}}].

\bibitem{abbott/2017b}
B.P.~Abbott, R.~Abbott, T.D.~Abbott, F.~Acernese, K.~Ackley, C.~Adams et~al.,
  \emph{Multi-messenger {{Observations}} of a {{Binary Neutron Star Merger}}},
  \href{https://doi.org/10.3847/2041-8213/aa91c9}{\emph{The Astrophysical
  Journal} {\bfseries 848} (2017) L12}.

\bibitem{radice/2018a}
D.~Radice, A.~Perego, F.~Zappa and S.~Bernuzzi, \emph{{{GW170817}}: Joint
  {{Constraint}} on the {{Neutron Star Equation}} of {{State}} from
  {{Multimessenger Observations}}},
  \href{https://doi.org/10.3847/2041-8213/aaa402}{\emph{The Astrophysical
  Journal Letters} {\bfseries 852} (2018) L29}.

\bibitem{motta/2019}
T.F.~Motta, A.M.~Kalaitzis, S.~Anti{\'c}, P.A.M.~Guichon, J.R.~Stone and
  A.W.~Thomas, \emph{Isovector {{Effects}} in {{Neutron Stars}}, {{Radii}}, and
  the {{GW170817 Constraint}}},
  \href{https://doi.org/10.3847/1538-4357/ab218e}{\emph{The Astrophysical
  Journal} {\bfseries 878} (2019) 159}.

\bibitem{margalit/2017}
B.~Margalit and B.D.~Metzger, \emph{Constraining the {{Maximum Mass}} of
  {{Neutron Stars}} from {{Multi-messenger Observations}} of {{GW170817}}},
  \href{https://doi.org/10.3847/2041-8213/aa991c}{\emph{The Astrophysical
  Journal Letters} {\bfseries 850} (2017) L19}.

\bibitem{bauswein/2017}
A.~Bauswein, O.~Just, H.-T.~Janka and N.~Stergioulas, \emph{Neutron-star
  {{Radius Constraints}} from {{GW170817}} and {{Future Detections}}},
  \href{https://doi.org/10.3847/2041-8213/aa9994}{\emph{The Astrophysical
  Journal} {\bfseries 850} (2017) L34}.

\bibitem{gamba/2019}
R.~Gamba, J.S.~Read and L.E.~Wade, \emph{The impact of the crust equation of
  state on the analysis of {{GW170817}}},
  \href{https://doi.org/10.1088/1361-6382/ab5ba4}{\emph{Classical and Quantum
  Gravity} {\bfseries 37} (2019) 025008}.

\bibitem{lourenco/2020}
O.~Louren{\c c}o, M.~Dutra, C.H.~Lenzi, S.K.~Biswal, M.~Bhuyan and
  D.P.~Menezes, \emph{Consistent {{Skyrme}} parametrizations constrained by
  {{GW170817}}},
  \href{https://doi.org/10.1140/epja/s10050-020-00040-z}{\emph{The European
  Physical Journal A} {\bfseries 56} (2020) 32}.

\bibitem{essick/2020}
R.~Essick, P.~Landry and D.E.~Holz, \emph{Nonparametric inference of neutron
  star composition, equation of state, and maximum mass with {{GW170817}}},
  \href{https://doi.org/10.1103/physrevd.101.063007}{\emph{Physical Review D}
  {\bfseries 101} (2020) 063007}.

\bibitem{lattimer/2001}
J.M.~Lattimer and M.~Prakash, \emph{Neutron {{Star Structure}} and the
  {{Equation}} of {{State}}}, \href{https://doi.org/10.1086/319702}{\emph{The
  Astrophysical Journal} {\bfseries 550} (2001) 426}.

\bibitem{lackey/2006}
B.D.~Lackey, M.~Nayyar and B.J.~Owen, \emph{Observational constraints on
  hyperons in neutron stars},
  \href{https://doi.org/10.1103/physrevd.73.024021}{\emph{Physical Review D}
  {\bfseries 73} (2006) 024021}.

\bibitem{bejger/2005}
M.~Bejger, T.~Bulik and P.~Haensel, \emph{Constraints on the dense matter
  equation of state from the measurements of {{PSR J0737-3039A}} moment of
  inertia and {{PSR J0751}}+1807 mass},
  \href{https://doi.org/10.1111/j.1365-2966.2005.09575.x}{\emph{Monthly Notices
  of the Royal Astronomical Society} {\bfseries 364} (2005) 635}.

\bibitem{ozel/2016}
F.~{\"O}zel and P.~Freire, \emph{Masses, {{Radii}}, and the {{Equation}} of
  {{State}} of {{Neutron Stars}}},
  \href{https://doi.org/10.1146/annurev-astro-081915-023322}{\emph{Annual
  Review of Astronomy and Astrophysics} {\bfseries 54} (2016) 401}.

\bibitem{Akmal-1998}
A.~Akmal, V.R.~Pandharipande and D.G.~Ravenhall, \emph{Equation of state of
  nucleon matter and neutron star structure},
  \href{https://doi.org/10.1103/PhysRevC.58.1804}{\emph{Phys. Rev. C}
  {\bfseries 58} (1998) 1804}.

\bibitem{Baldo-1997}
M.~Baldo, I.~Bombaci and G.F.~Burgio, \emph{{Microscopic nuclear equation of
  state with three-body forces and neutron star structure}}, {\emph{Astron.
  Astrophys.} {\bfseries 328} (1997) 274}
  [\href{https://arxiv.org/abs/astro-ph/9707277}{{\ttfamily
  astro-ph/9707277}}].

\bibitem{Friedman-1981}
B.~Friedman and V.~Pandharipande, \emph{Hot and cold, nuclear and neutron
  matter},
  \href{https://doi.org/https://doi.org/10.1016/0375-9474(81)90649-7}{\emph{Nuclear
  Physics A} {\bfseries 361} (1981) 502}.

\bibitem{Douchin-2001}
{Douchin, F.} and {Haensel, P.}, \emph{A unified equation of state of dense
  matter and neutron star structure},
  \href{https://doi.org/10.1051/0004-6361:20011402}{\emph{A\&A} {\bfseries 380}
  (2001) 151}.

\bibitem{Prakash-1988}
M.~Prakash, T.L.~Ainsworth and J.M.~Lattimer, \emph{Equation of state and the
  maximum mass of neutron stars},
  \href{https://doi.org/10.1103/PhysRevLett.61.2518}{\emph{Phys. Rev. Lett.}
  {\bfseries 61} (1988) 2518}.

\bibitem{Wiringa-1988}
R.B.~Wiringa, V.~Fiks and A.~Fabrocini, \emph{Equation of state for dense
  nucleon matter}, \href{https://doi.org/10.1103/PhysRevC.38.1010}{\emph{Phys.
  Rev. C} {\bfseries 38} (1988) 1010}.

\bibitem{Engvik-1996}
L.~Engvik, E.~Osnes, M.~Hjorth-Jensen, G.~Bao and E.~Ostgaard, \emph{Asymmetric
  nuclear matter and neutron star properties},
  \href{https://doi.org/10.1086/177827}{\emph{The Astrophysical Journal}
  {\bfseries 469} (1996) 794}.

\bibitem{Muther-1987}
H.~Müther, M.~Prakash and T.~Ainsworth, \emph{The nuclear symmetry energy in
  relativistic brueckner-hartree-fock calculations},
  \href{https://doi.org/https://doi.org/10.1016/0370-2693(87)91611-X}{\emph{Physics
  Letters B} {\bfseries 199} (1987) 469}.

\bibitem{Muller-1996}
H.~Müller and B.D.~Serot, \emph{Relativistic mean-field theory and the
  high-density nuclear equation of state},
  \href{https://doi.org/https://doi.org/10.1016/0375-9474(96)00187-X}{\emph{Nuclear
  Physics A} {\bfseries 606} (1996) 508}.

\bibitem{Zuo-1999}
W.~Zuo, I.~Bombaci and U.~Lombardo, \emph{Asymmetric nuclear matter from an
  extended brueckner-hartree-fock approach},
  \href{https://doi.org/10.1103/PhysRevC.60.024605}{\emph{Phys. Rev. C}
  {\bfseries 60} (1999) 024605}.

\bibitem{Pandharipande-1975}
V.~Pandharipande and R.~Smith, \emph{A model neutron solid with $\pi0$
  condensate},
  \href{https://doi.org/https://doi.org/10.1016/0375-9474(75)90415-7}{\emph{Nuclear
  Physics A} {\bfseries 237} (1975) 507}.

\bibitem{Glendenning-1999}
N.K.~Glendenning and J.~Schaffner-Bielich, \emph{First order kaon condensate},
  \href{https://doi.org/10.1103/PhysRevC.60.025803}{\emph{Phys. Rev. C}
  {\bfseries 60} (1999) 025803}.

\bibitem{Glendenning-1985}
N.K.~{Glendenning}, \emph{{Neutron stars are giant hypernuclei ?}},
  \href{https://doi.org/10.1086/163253}{\emph{Astrophys J} {\bfseries 293}
  (1985) 470}.

\bibitem{Lackey-2006}
B.D.~Lackey, M.~Nayyar and B.J.~Owen, \emph{Observational constraints on
  hyperons in neutron stars},
  \href{https://doi.org/10.1103/PhysRevD.73.024021}{\emph{Phys. Rev. D}
  {\bfseries 73} (2006) 024021}.

\bibitem{Prakash-1995}
M.~Prakash, J.R.~Cooke and J.M.~Lattimer, \emph{Quark-hadron phase transition
  in protoneutron stars},
  \href{https://doi.org/10.1103/PhysRevD.52.661}{\emph{Phys. Rev. D} {\bfseries
  52} (1995) 661}.

\bibitem{Alford-2005}
M.~Alford, M.~Braby, M.~Paris and S.~Reddy, \emph{Hybrid stars that masquerade
  as neutron stars}, \href{https://doi.org/10.1086/430902}{\emph{The
  Astrophysical Journal} {\bfseries 629} (2005) 969}.

\bibitem{jolliffe/2013}
I.T.~Jolliffe, \emph{Principal {{Component Analysis}}}, {Springer Science \&
  Business Media} (2013).

\bibitem{cox/2008}
M.A.A.~Cox and T.F.~Cox, \emph{Multidimensional {{Scaling}}},
  \href{https://doi.org/10.1007/978-3-540-33037-0_14}{\emph{Handbook of Data
  Visualization} (2008) 315}.

\bibitem{NIPS2002_6150ccc6}
G.E.~Hinton and S.~Roweis, \emph{Stochastic neighbor embedding},  in
  \emph{Advances in Neural Information Processing Systems}, S.~Becker, S.~Thrun
  and K.~Obermayer, eds., vol.~15, {MIT Press}, 2002.

\end{thebibliography}\endgroup
\end{document}